# Pulse-to-pulse spectral phase characterization of mid-infrared pulses at megahertz rates

Zhihao Deng[1,†], Zicong Xu[2,3,†], Kazuki Hashimoto[2], Gabriel Demontigny[4,5], Denis V. Seletskiy[4,5], Takuro Ideguchi[1,2,3,6*]

[1] Department of Physics, The University of Tokyo, Tokyo, Japan
[2] Institute for Photon Science and Technology, The University of Tokyo, Tokyo, Japan
[3] Department of Advanced Materials Science, The University of Tokyo, Chiba, Japan
[4] femtoQ Laboratory, Department of Engineering Physics, Polytechnique Montréal, Québec, Canada
[5] femtoQ Laboratory, Department of Physics and Astronomy, University of New Mexico, NM, USA
[6] RIKEN Center for Advanced Photonics, RIKEN, Saitama, Japan
[†] These authors contributed equally to this work.
[*]ideguchi@edu.k.u-tokyo.ac.jp

## Abstract

Pulse-resolved spectral phase measurement of mid-infrared (MIR) pulses is essential for many applications, from precise waveform control to ultrafast quantum optics. However, conventional MIR pulse characterization techniques are typically limited to sub-kHz-rate operation, leaving a substantial speed mismatch with MIR sources operating at kHz or MHz rates. Here, we introduce time-stretch upconversion-based mid-infrared pulse evaluation (TSUBAME), a technique that enables pulse-to-pulse spectral phase characterization of ultrashort MIR pulses at the laser repetition rate. TSUBAME combines MIR-to-NIR (near-infrared) upconversion, time-stretch, and spectral interferometry to achieve scan-free high-speed spectral phase measurements. We validated the technique by measuring MIR pulses spanning 4.98-5.30 μm while introducing well-defined dispersion, obtaining excellent agreement with theoretical predictions. Operating at a measurement rate of 1 MHz, TSUBAME achieves the fastest single-pulse-resolved spectral phase characterization of MIR pulses reported to date. As a further demonstration, we captured dynamic spectral phase variations on a microsecond timescale. TSUBAME provides a powerful tool for real-time monitoring and optimization of high-repetition-rate MIR pulses, with potential applications in strong-field physics, high-harmonic generation, and coherent molecular control.

## Introduction

Measuring the spectral phase of ultrashort mid-infrared (MIR) laser pulses is fundamental to a wide range of applications in which the temporal waveform plays an essential role. In strong-field physics and high-harmonic generation (HHG), the MIR waveform governs nonlinear light-matter interactions and the resulting emitted fields[1]. Precise control of the MIR spectral phase is also essential for few-cycle waveform synthesis[2], coherent molecular control[3], and vibrational ladder climbing[4], where tailored optical fields are used to drive specific light-matter interactions. Because these applications are sensitive to the waveform of each individual pulse, pulse-to-pulse spectral phase variations can directly affect the underlying light-matter interaction. Such variations may originate from pump fluctuations[5] in wavelength-conversion-based MIR sources[6,7] and intracavity dynamics[8] in mode-locked MIR lasers[9–11], or more fundamentally, due to the quantum fluctuations[12–14]. In all cases, however, they are obscured in averaged measurements. Therefore, pulse-resolved MIR spectral phase measurement at the native repetition rate of the source, typically in the kHz or MHz range, is essential for diagnosing source instabilities, controlling pulse waveforms, and exploiting ultrafast MIR pulses in high-speed applications.

Various techniques have been developed for measuring the spectral phase of ultrafast MIR pulses[15], including frequency-resolved optical gating (FROG)[16,17], dispersion scan (d-scan)[18,19], spectral phase interferometry for direct electric-field reconstruction (SPIDER)[20,21], and electro-optic sampling (EOS)[22–24]. However, extending these techniques to pulse-resolved spectral phase measurements has been challenging. Conventional implementations of FROG, d-scan, and EOS rely on mechanical scanning of the pulse delay or wedge-pair separation, fundamentally limiting their acquisition speed and making them unsuitable for capturing spectral phase evolution from one pulse to the next. Scanning-free approaches such as SPIDER offer the potential for high-rate pulse characterization, but in practice their acquisition rates are typically restricted to the millisecond timescale by the readout speed of spectrometer sensor arrays[20,21]. Consequently, pulse-to-pulse spectral phase measurements of MIR pulses at kHz and MHz repetition rates have remained inaccessible.

Time-stretch techniques, also known as dispersive Fourier-transform techniques, provide a powerful strategy for MHz-rate pulse-resolved spectral measurements by converting optical spectra into temporal intensity waveforms at the native repetition rate of ultrafast sources[25,26]. Time-stretch spectroscopy has been extensively developed in the near-infrared (NIR) telecom region, where high-quality pulse stretching and detection can be readily implemented through low-loss fiber propagation and high-bandwidth photodetection. In contrast, applying time-stretch techniques in the MIR region has long remained challenging because of the lack of suitable dispersive media and photodetection techniques. Recently, several approaches have been proposed, including the use of a free-space angular-chirp-enhanced delay (FACED)[27] and MIR-to-NIR upconversion followed by time stretching in NIR fibers[28,29]. However, these approaches have been limited to spectral intensity measurements, leaving direct spectral phase characterization of ultrashort MIR pulses unexplored.

In this study, we demonstrate time-stretch upconversion-based mid-infrared pulse evaluation (TSUBAME, Japanese for the bird “swallow”, an homage to fast, gracious and highly agile bird chirping mid-flight), which enables pulse-to-pulse spectral phase characterization of ultrashort MIR pulses at the native repetition rate of the source. The central concept of TSUBAME is to combine time-stretch detection with spectral interferometry. For time-stretch detection of MIR pulses, TSUBAME adopts an upconversion-based approach to achieve the high spectral resolution required for spectral phase characterization[28,29]. For spectral interferometry, the upconverted pulse interferes with a reference NIR pulse with a known spectral phase, thereby encoding phase information into the spectral intensity and enabling direct spectral phase retrieval through a non-iterative decoding algorithm. Using TSUBAME, we characterized broadband MIR pulses spanning 4.98-5.30 μm (1887-2008 $cm^{-1}$) with a spectral resolution of 8.7 nm (3.26 $cm^{-1}$) at a measurement rate of 1 MHz. The achieved spectral phase precision was 3.73-22.16 mrad. The performance of the spectral phase measurements was validated by introducing known chirp to the MIR pulses using materials with well-defined group-delay dispersion (GDD). As a proof-of-principle demonstration of rapid dynamics capture, we measured microsecond-scale spectral phase variations across successive MIR pulses. We anticipate that TSUBAME will provide a useful tool for monitoring MIR pulse dynamics and enabling real-time manipulation of spectral phase in applications such as strong-field physics, HHG, coherent molecular control, and ultrafast quantum optics.

## Results

### Working principle of TSUBAME

Figure 1 illustrates the general working principle of TSUBAME. An ultrashort MIR pulse under test is combined with a narrowband picosecond NIR pulse (narrowband TSUBAME pulse) for upconversion via a nonlinear optical process such as sum-frequency generation (SFG). When the bandwidth of the narrowband TSUBAME pulse is sufficiently narrow compared to the MIR pulse, the spectral intensity and phase of the MIR pulse are mapped onto the upconverted NIR pulse. A broadband femtosecond NIR pulse (broadband TSUBAME pulse) is then prepared to spectrally overlap with the upconverted pulse. After combining the broadband TSUBAME pulse and the upconverted pulse at a beamsplitter (BS), spectral interference is observed, with fringe spacing determined by the relative delay between the two pulses. In addition, the spectral phase difference between the two pulses introduces a gradual variation in the fringe spacing across the spectrum. If the spectral phase of the broadband TSUBAME pulse is known in advance, that of the MIR pulse can be retrieved from the interference fringes. Further theoretical details are provided in Supplementary Note 1.

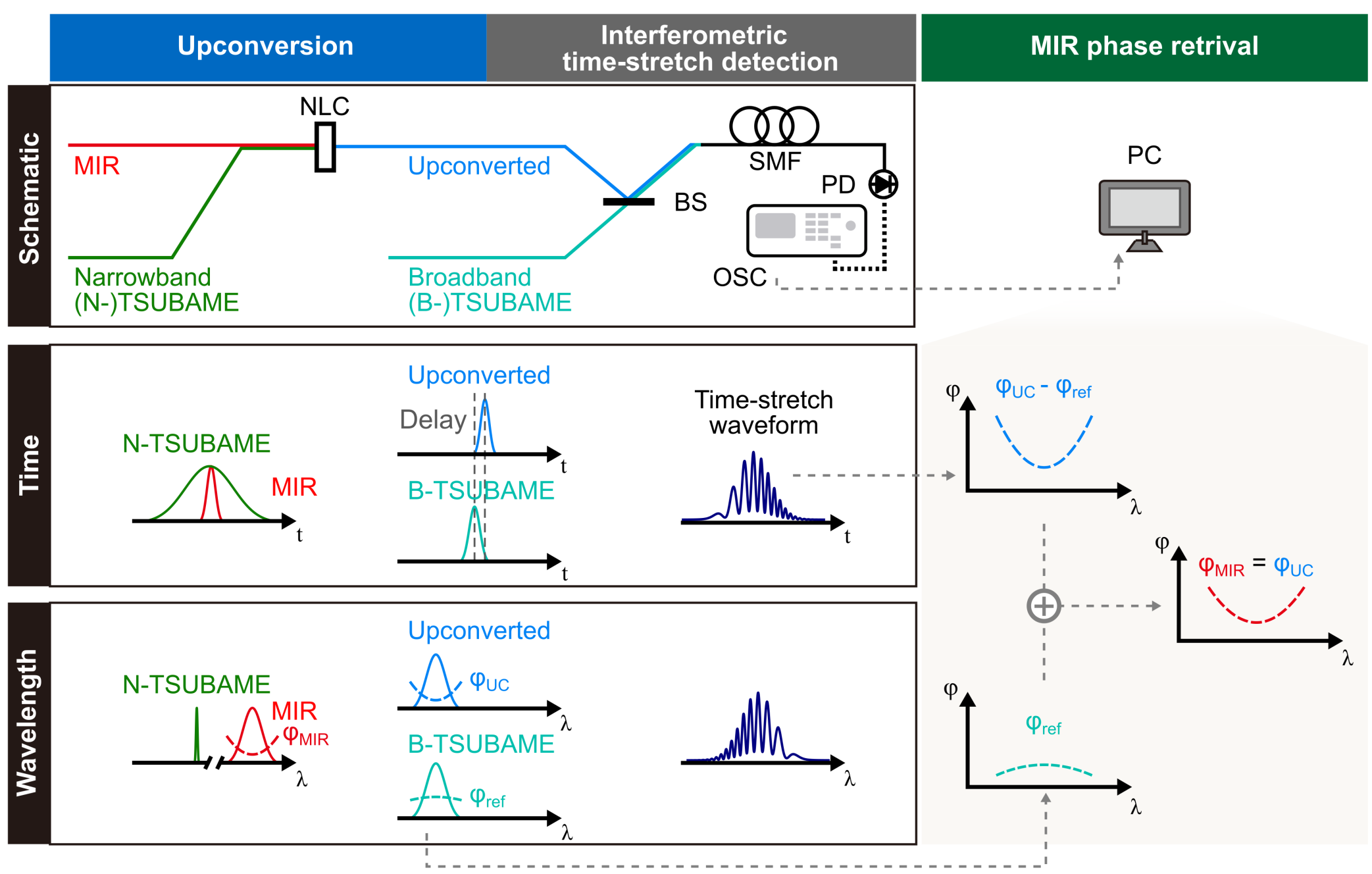

**Figure 1**. **Working principle of TSUBAME.** The spectral phase of the MIR pulse is retrieved using TSUBAME pulses consisting of a narrowband and a broadband pulse. The MIR pulse is upconverted to the NIR region by the narrowband TSUBAME pulse, preserving its spectral intensity and phase. The upconverted pulse interferes with the broadband TSUBAME pulse, and the resulting spectral interference is recorded in the time domain on a pulse-to-pulse basis using time-stretch detection. The spectral phase difference between the upconverted and broadband TSUBAME pulses is extracted from the time-stretch waveform via time-to-wavelength mapping. By adding the independently measured spectral phase of the broadband TSUBAME pulse, the spectral phase of the MIR pulse is retrieved. NLC: nonlinear crystal, BS: beamsplitter, SMF: single-mode fiber, PD: photodetector, OSC: oscilloscope. $t$: time, $\lambda$: wavelength, $\varphi$: spectral phase.

To enable pulse-resolved detection of the spectral interference, the two pulses are coupled into a long single-mode fiber (SMF). The fiber is sufficiently long to map the spectral profile of each pulse into the temporal domain via GDD, analogous to a Fourier transform in the large-dispersion limit. After time-stretching, the spectral fringes encoded in the temporal intensity waveforms are detected by a high-speed photodetector (PD) and recorded using a high-speed oscilloscope (OSC), achieving high spectral resolution. Through post-processing of the fringe-encoded temporal waveforms, the spectral phase of the MIR pulse can be retrieved after calibration using the pre-characterized spectral phase of the broadband TSUBAME pulse. Meanwhile, the spectral intensity of the MIR pulse can be obtained from a simultaneously acquired time-stretched upconversion spectrum, with appropriate wavelength conversion from NIR back to MIR. Therefore, TSUBAME enables access to both the spectral intensity and phase of ultrashort MIR pulses on a pulse-to-pulse basis, with measurement speed limited only by the laser repetition rate. In this work, however, we primarily focus on spectral phase retrieval using TSUBAME.

**Experimental implementation of TSUBAME**

Figure 2 depicts the schematic of the experimental setup implementing TSUBAME. For convenience, both the MIR pulses and the TSUBAME pulses were derived from a common laser source. As the light source, we used a Yb-doped fiber laser centered at 1037 nm, delivering femtosecond pulses at a repetition of 50 MHz (YLMO HP, Menlo Systems). At the laser output, an acousto-optic modulator (AOM; 1064AOM-2, AeroDIODE) was employed as a pulse picker to reduce the repetition rate to 1 MHz. This significantly improves the stability of the subsequent spectral broadening stage, which is essential for generating both the broadband TSUBAME pulse and the MIR pulse. The output from the polarization-maintaining (PM) fiber-coupled AOM was connected to a 90/10 PM fiber coupler (CP-PS-1X2-1030-10/90-0-5-3X54-1-KK, Kokyo). The 90% port was used to generate the ultrashort MIR pulse and the broadband TSUBAME pulse, while the 10% port was used to prepare the narrowband TSUBAME pulse.

Prior to MIR generation, the pulses were amplified using a homemade PM double-clad Yb-doped fiber amplifier (YDFA). After compression to the transform limit (230 fs) using a grating pair, the pulses were coupled into a 55-mm-long SMF (1060XP, Thorlabs). The high pulse energy (~48.3 nJ) and the small core area of the SMF induced strong self-phase modulation (SPM), resulting in substantial spectral broadening. Following compression with a pair of chirped mirrors (PC1611, Ultrafast Innovations), pulses as short as 9.7 fs were obtained. A detailed description of the spectral broadening system can be found in our previous work[30]. The resultant 9.7-fs pulse was split using a 5/95 BS. The 5% output served as the broadband TSUBAME pulse, while the 95% output was used to generate MIR pulses via intra-pulse difference-frequency generation (IPDFG) in a 1-mm-long fan-out periodically poled lithium niobate (PPLN) crystal (AC-Y161201-37-08-02, HC Photonics) operated at 120 °C. The fan-out grating period was set to 22.8 μm to satisfy the phase-matching condition, yielding an MIR output power of 40 μW with a 48.3 mW NIR pump. The generated MIR spectrum was centered at 5.15 μm with a full width at half maximum (FWHM) of 0.190 μm. The corresponding transform-limited pulse duration (FWHM) was estimated to be 205 fs under the Gaussian approximation. Residual NIR pump light after the IPDFG process was removed using a germanium (Ge) window with an AR coating for the 3-6 μm wavelength range, which acted as a long-pass filter.

For spectral phase measurement of the MIR pulse, a narrowband TSUBAME pulse was used to upconvert the MIR pulse to the NIR region while preserving its spectral phase information (see Supplementary Note 1 for theoretical derivation). The narrowband TSUBAME pulse was derived from the 10% output port of the fiber coupler located after the pulse picker (not shown in Fig. 2). The broadband pulse was spatially dispersed using a reflective diffraction grating (1200 lines/mm), and a fiber collimator was positioned at a sufficient distance from the grating such that only a narrow spectral component could be coupled. To amplify the selected narrowband pulse, we constructed a two-stage PM-YDFA, comprising a single-clad gain fiber in the first stage and a double-clad gain fiber in the second stage. At the YDFA output, the narrowband TSUBAME pulse had a central wavelength of 1032.8 nm with an FWHM of 0.27 nm (see Fig. S1(c) in Supplementary Note 2). From this bandwidth, the transform-limited pulse duration was estimated to be 5.8 ps.

For the upconversion process, SFG was performed between the MIR pulse and the narrowband TSUBAME pulse in a 1-mm-long fan-out PPLN crystal (AC-Y161201-37-08-02, HC Photonics) operated at 120 °C. The poling period was tuned to 21 µm. The upconverted pulse spanned a wavelength range from 855.4 to 865.4 nm, with an average power of 21.5 µW, obtained from a 40 µW MIR pulse and a 446.2 mW narrowband TSUBAME pulse. After SFG, the upconverted pulse was collinearly combined with the broadband TSUBAME pulse using a 50/50 cube BS (BS014, Thorlabs). The bandwidth of the broadband TSUBAME pulse guarantees partial overlap with the upconverted pulse, enabling spectral interference. A cube BS was used instead of a plate BS to ensure that both input pulses experienced similar dispersion, therefore minimizing its effects on the spectral interference. One output port of the BS was directed to the time-stretch detection system, while the other (not shown in Fig. 2) was used to monitor the repetition signal and provide a trigger for the measurement. For selective time-stretch detection of the interfering spectral components, a band-pass filter (BPF), consisting of an 800 nm long-pass filter (FELH0800, Thorlabs) and a 900 nm short-pass filter (FESH0900, Thorlabs), was placed after the BS.

The spectral interference between the upconverted pulse and the broadband TSUBAME pulse was detected using time-stretching to resolve the pulse-to-pulse spectral phase. Time-stretching was achieved by coupling the beam centered at 860.4 nm into a Corning SMF-28 fiber commonly used in optical communications. The fiber exhibits a dispersion[31] of −84 ps/(nm·km) and a propagation loss[32] of 1.75 dB/km at 860.4 nm. After propagation through 4 km of fiber, the pulse was stretched to a temporal duration of approximately 20 ns, with its spectral information mapped onto the time axis. Note that SMF-28 supports multimode propagation at the operating wavelength, which can disrupt the one-to-one wavelength-to-time mapping because of modal dispersion. To suppress higher-order spatial modes, the final section of the fiber was bent with a radius of approximately 1 cm. To resolve the fine spectral fringes, a 10 GHz InGaAs PD (RXM10AF, Thorlabs) was used, followed by a 16 GHz oscilloscope (WaveMaster 816Zi-B, Teledyne LeCroy) operating at 80 GSamples/s with 8-bit vertical resolution. The average optical power at the PD input was measured to be 1.35 µW.

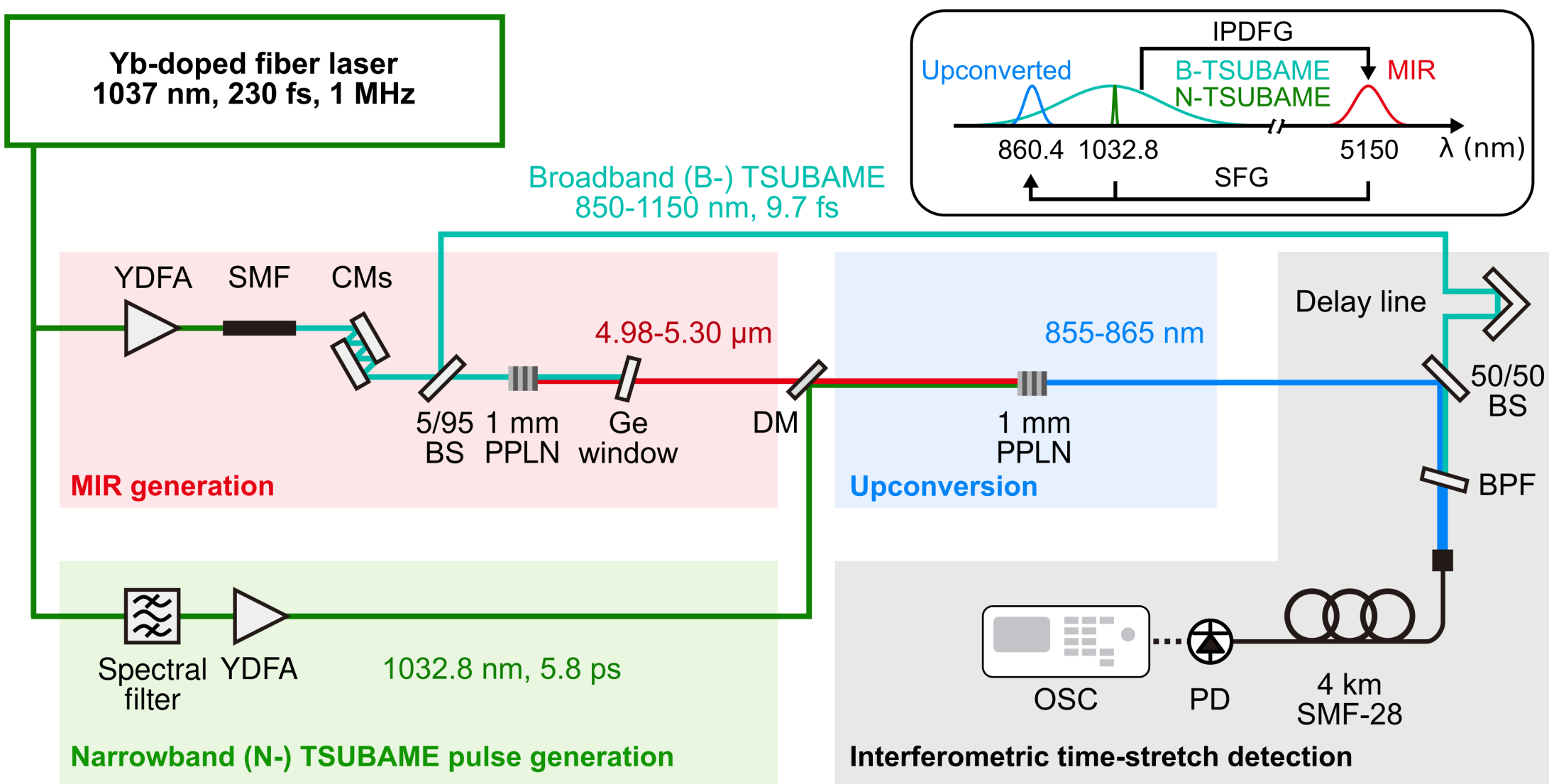

**Figure 2**. **Schematic of the TSUBAME system.** MIR and TSUBAME pulses are derived from the same laser source. The MIR pulse is generated via IPDFG (intra-pulse difference-frequency generation), the broadband TSUBAME pulse via spectral broadening, and the narrowband TSUBAME pulse via spectral filtering. The MIR pulse and the narrowband TSUBAME pulse are used for upconversion

(sum-frequency generation, SFG), producing an NIR pulse that interferes with the broadband TSUBAME pulse. The resulting interference is detected using a time-stretch detection system. The upper right inset shows a spectral diagram of wavelength conversion processes in the TSUBAME system. YDFA: ytterbium (Yb)-doped fiber amplifier, SMF: single-mode fiber, CMs: chirped mirrors, BS: beamsplitter, PPLN: periodically poled lithium niobate, DM: dichroic mirror, BPF: band-pass filter, PD: photodetector, OSC: oscilloscope.

**Pulse-to-pulse MIR spectral phase retrieval**

First, we consecutively acquired 400 time-stretch waveforms at a rate of 1 MHz. Figure 3(a) shows three sequential waveforms corresponding to independently acquired upconverted pulses and broadband TSUBAME pulses. The raw data were recorded in the time domain using an OSC, and a calibrated time-to-wavelength mapping (see Supplementary Note 2) was applied to display the data on a wavelength scale. Using the waveforms of the upconverted pulses, we evaluated the single-shot signal-to-noise ratio (SNR) of the time-stretch detection. Each raw waveform was first smoothed using a Savitzky-Golay filter, and the SNR was then defined as the ratio of the peak intensity to the standard deviation calculated within a selected region around the signal peak. Figure 3(d) shows the variation in single-shot SNR across 18 consecutive waveforms as a function of the average optical power of upconverted pulses incident on the PD. The SNR increases approximately linearly with optical power, indicating that the dominant noise sources arise from the PD and the OSC. The maximum single-shot SNR of 65 is achieved at an average power of 1.66 μW, which is limited by the saturation threshold of the PD.

When both the upconverted pulse and the broadband TSUBAME pulse were present, their spectral overlap produced interference fringes, as shown in Fig. 3(b). The waveforms were Fourier transformed after application of a Kaiser window function. A representative frequency domain spectrum is shown in Fig. 3(e). As the fringe spacing is determined by the delay in the broadband TSUBAME path, we carefully adjusted it so that the fringe frequency (6.7 GHz) remains within the PD bandwidth (10 GHz). Since the spectral phase information is encoded in the interference fringes, we applied a bidirectional digital BPF with a passband of 6.1-7.3 GHz to isolate the signal centered at 6.7 GHz while suppressing noise, as shown in Fig. 3(e).

After band-pass filtering, a Hilbert transform was applied to extract the phase of the fringes, as shown in Fig. 3(c). The use of a bidirectional BPF ensures that no phase distortion is introduced during filtering. The sign of the retrieved spectral phase was determined from the temporal ordering of the upconverted pulse and the broadband TSUBAME pulse at the detector. In addition, we evaluated the single-shot phase precision by dividing the background noise by the fringe envelope after applying the BPF (see Supplementary Note 3 for more details). A representative single-shot phase precision curve is shown in Fig. 3(c) for the pulse centered at 999 ns, yielding a minimum phase error of 3.73 mrad at the envelope peak and 22.16 mrad at the edges (at the $1/e^2$ level). The highest phase precision at the envelope peak is limited by both the PD saturation and detector noise coming from the PD and OSC.

Using the acquired data, we retrieved both the spectral intensity and spectral phase of the MIR pulse, as shown in Fig. 3(f). The upconverted spectrum provides the MIR spectral intensity information, while its interference with the broadband TSUBAME pulse encodes the MIR spectral phase. In the upconversion process from MIR to NIR, the use of a narrowband TSUBAME pulse as the pump ensures a one-to-one wavelength mapping between the MIR and NIR domains while preserving the spectral intensity and phase information (see Supplementary Note 1 for a detailed theoretical derivation). Accordingly, the measured NIR upconverted spectrum was directly converted to the MIR wavelength scale using the central wavelength of narrowband TSUBAME pulse.

For spectral phase retrieval, the directly measured quantity is the phase difference between the upconverted pulse and the broadband TSUBAME pulse. The spectral phase of the broadband TSUBAME pulse, before it was combined with the upconverted pulse at the BS, was independently characterized using a commercial NIR SPIDER (Venteon SPIDER, Laser Quantum, now Novanta), as shown in Fig. 3(f) (dashed black curve). By adding this reference phase to the phase extracted from the spectral fringes, we could obtain the spectral phase of the upconverted pulse, which corresponds to that of the MIR pulse. A representative result of MIR spectral phase (solid black curve) is shown in Fig. 3(f) for the pulse centered at 999 ns. The parabolic profile of the retrieved MIR spectral phase indicates the presence of GDD. By fitting the phase with a second-order polynomial, we estimated a GDD of $-14125$ fs$^2$, which agrees well with the theoretical value of $-14230$ fs$^2$, calculated by accounting for the dispersive contributions of the optical components in the system.

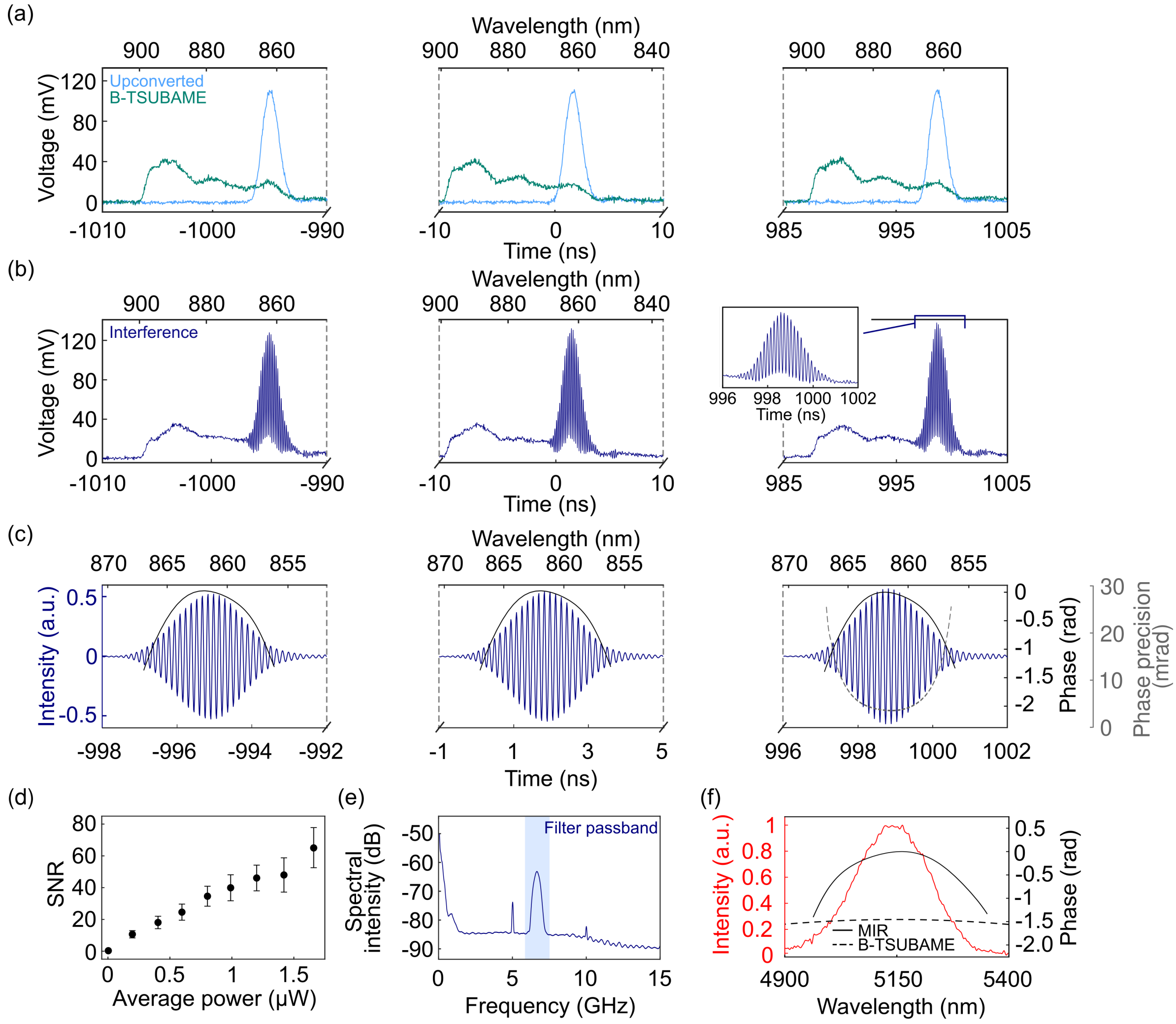

**Figure 3. Procedures for retrieving the MIR spectral phase and characterizations of the TSUBAME system.** (a) A sequence of three consecutively measured time-stretch waveforms (spectra) of the upconverted pulse (light blue) and the broadband TSUBAME pulse (teal) at a repetition rate of 1 MHz. (b) Time-stretch waveforms (spectra) of the interference between the upconverted and broadband TSUBAME pulses. The inset shows a zoomed view of the spectral fringes. (c) Spectral fringes after filtering (solid navy), the corresponding spectral phase (solid black), and single-shot phase precision (dashed gray). (d) Single-shot SNR of the upconverted pulse. (e) Fourier-transform of interferometric time-stretch waveforms. (f) A representative retrieved MIR intensity spectrum (solid red) and spectral phase (solid black), together with the spectral phase of the broadband TSUBAME pulse (dashed black). These results correspond

to the pulse centered at 999 ns.

**Validation with static $CaF_2$ plates**

To validate the capability of our system for MIR spectral phase evaluation, we introduced additional dispersion to the pulse under test by inserting $CaF_2$ plates of varying thickness. The $CaF_2$ plates (WG51050, Thorlabs; #11-874, Edmund Optics) are transparent at the operating wavelength while introducing well-defined dispersion. The plates were placed after the MIR beam and the narrowband TSUBAME were combined, ensuring that both paths experienced a similar optical path length. The delay in the broadband TSUBAME arm was adjusted accordingly.

Figure 4(a) shows the retrieved MIR spectral phase for $CaF_2$ thicknesses of 0, 3, 5, and 8 mm. The 0 mm case corresponds to the absence of an inserted plate, while the 8 mm case was realized by inserting the 3 mm and 5 mm plates together. For each thickness, 400 consecutive pulses were recorded. As shown in Fig. 4 (a), the quadratic spectral phase becomes progressively steeper with increasing $CaF_2$ thickness, indicating increased dispersion (anomalous dispersion at operating wavelength). The zeroth- and first- order terms were removed from the phase to emphasize the GDD.

For each pulse, the spectral phase was fitted with a second-order polynomial, and the GDD was extracted from the quadratic coefficient. The mean and standard deviation of the GDD values for all 400 pulses at each thickness are summarized in Fig. 4(b). The observed variance in GDD arises partly from the limited phase precision of the measurement but is dominated by pulse-to-pulse fluctuations in the spectral phase of the MIR pulses generated via IPDFG (see Discussions for more details). A linear fit to the data yields a group-velocity dispersion (GVD) of $-738.2 \pm 73.3$ $fs^2$/mm. This value agrees well with the literature value for $CaF_2$ ($-699.3$ $fs^2$/mm at 5.15 μm)[33], confirming the validity of the TSUBAME measurement.

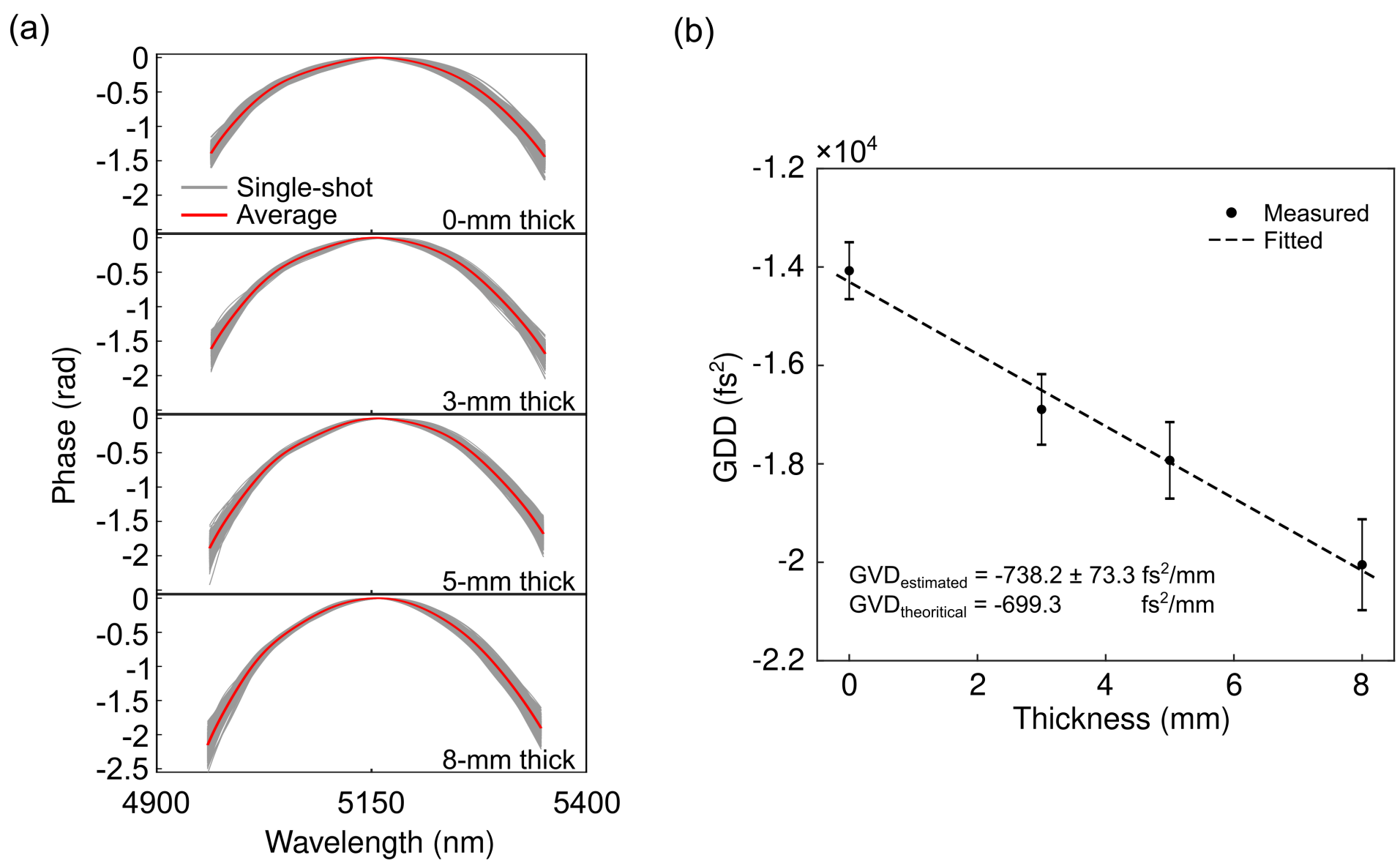

**Figure 4. Static GDD measurements of $CaF_2$ plates with different thicknesses at 1 MHz.** (a) Overlay of retrieved MIR spectral phases (gray) from 400 consecutive pulses, along with their average (red). (b) Retrieved GDD values from consecutive pulses for different $CaF_2$ thicknesses.

**Dynamic spectral-phase change measurements**

Next, we measured the spectral phase evolution of MIR pulses using TSUBAME to demonstrate its capability for capturing rapidly varying spectral phases. A 3-mm-thick $CaF_2$ plate mounted on a translation rail was manually translated to switch the presence of the plate in the MIR beam path on a sub-millisecond timescale. To capture the full switching dynamics, the acquisition time window of the OSC was extended. However, maintaining a high sampling rate (80 GSamples/s) for high spectral resolution limits the available OSC memory for long acquisitions. To address this, we recorded one waveform out of every ten pulses using the hold-off function, thereby reducing memory usage. Although this reduces the effective temporal resolution to 10 μs, it remains sufficient for this demonstration. For higher temporal resolution over longer acquisition windows, a digitizer with larger memory capacity would be advantageous.

The $CaF_2$ plate was placed in the common path of the MIR pulse and the narrowband TSUBAME pulse before SFG, thereby ensuring similar changes in optical path length and minimizing variations in SFG efficiency. Nevertheless, the insertion of the plate introduces an additional relative delay with respect to the broadband TSUBAME pulse, which cannot be precisely synchronized with the manual switching process. As a result, the interference fringe frequency differs depending on whether the MIR beam propagates through air or the $CaF_2$ plate. In this measurement, the delay in the broadband TSUBAME arm was carefully adjusted so that the fringe frequency remained within the PD bandwidth in both cases.

Figure 5(a) shows the RF spectrogram obtained from the Fourier transform of the time-domain fringe waveforms over a 20 ms acquisition window. A bidirectional digital high-pass filter with a cutoff frequency of 1 GHz was applied to suppress non-interferometric components and noise. In the 0-2 ms time window, the MIR beam propagates in air, resulting in a fringe frequency centered at 2.4 GHz. In the 18-20 ms window, the MIR beam passes through the 3-mm $CaF_2$ plate, and the fringe frequency shifts to 7.8 GHz. In the intermediate region (4-6 ms), a component centered at 5.4 GHz is observed, corresponding to the frequency difference between the two cases. This feature indicates that the MIR beam partially overlaps the edge of the $CaF_2$ plate, arising from interference between portions of the beam propagating through air and $CaF_2$. Finally, we analyzed 2000 waveforms and extracted the corresponding GDD values for each pulse. As shown in Fig. 5(b), the GDD values in the 0-2 ms and 18-20 ms windows are centered at $-14130$ and $-16961$ fs$^2$, respectively. These values are consistent with those obtained in the static measurements for 0 and 3 mm $CaF_2$, confirming the capability of TSUBAME for precise spectral phase measurement under dynamically varying conditions.

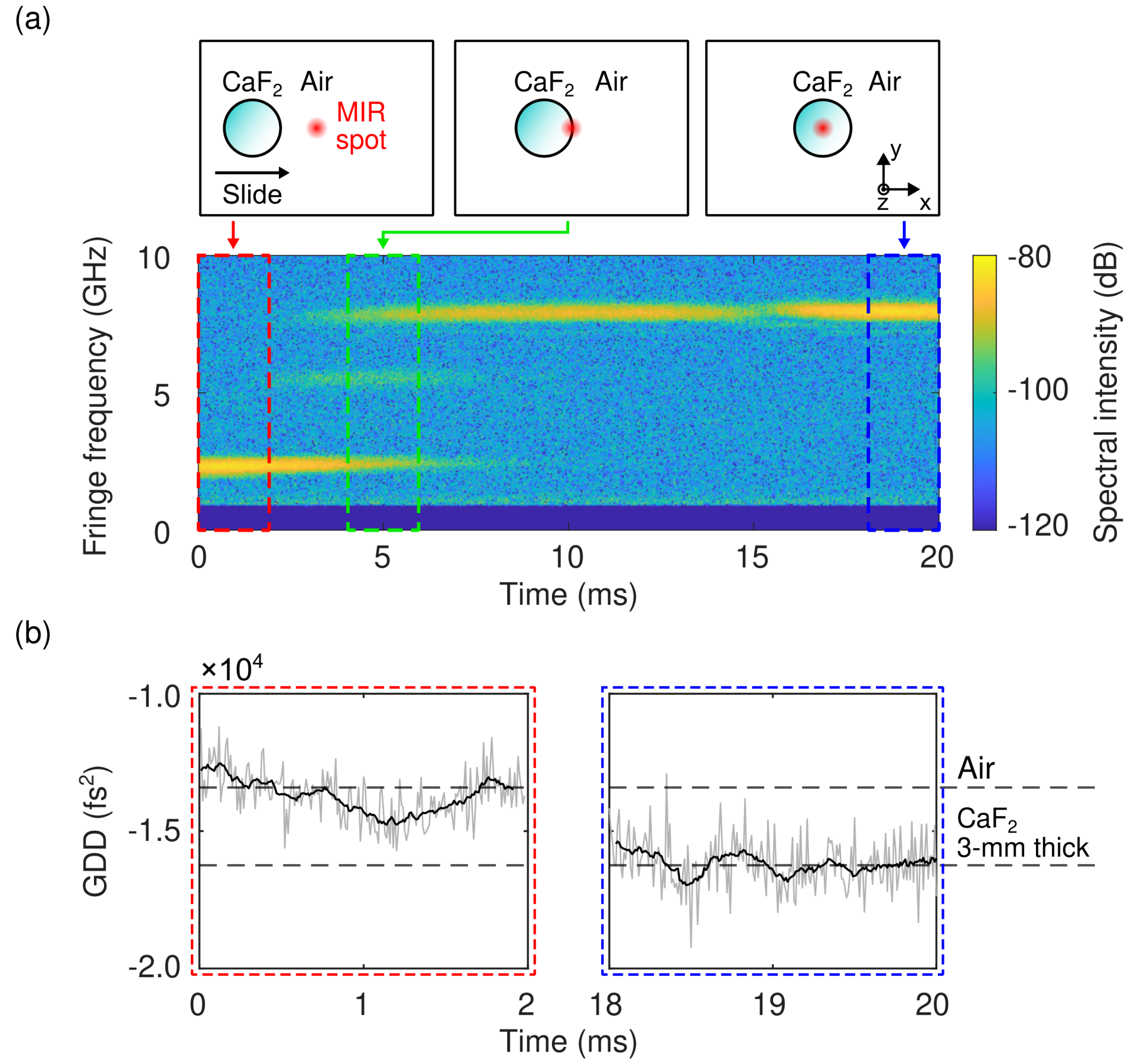

**Figure 5. Dynamic GDD measurements using TSUBAME.** (a) RF spectrogram of interferometric fringes measured at 100 kSpectra/s over 20 ms, obtained by rapidly switching the presence of a 3-mm-thick $CaF_2$ plate. The $CaF_2$ plate was mounted on a rail and translated along the x-direction, while the MIR beam propagated along the z-direction. (b) Time-resolved GDD values at 100 kSpectra/s (solid gray) in the 0-2 ms and 18-20 ms regions (corresponding to the red and blue dashed boxes in (a), respectively), along with a 15-pulse moving average (solid black). The dashed gray lines indicate the GDD values for 0- and 3-mm $CaF_2$ plates evaluated in Fig. 4(b).

## Discussion

### Performance of the current TSUBAME implementation

As a spectral phase characterization technique, TSUBAME achieves a phase precision of up to 3.73 mrad at 5.15 µm in the present setup, corresponding to a GDD uncertainty of 3.32 $fs^2$ and a third-order dispersion (TOD) uncertainty of 221 $fs^3$. This precision is more than sufficient for the current demonstrations using MIR pulses with a transform-limited pulse duration of 205 fs. The phase precision could be further improved by employing a PD with a higher saturation threshold or lower noise-equivalent power, together with a digitizer offering higher vertical resolution (*e.g.*, >12 bits). Notably, the static $CaF_2$ plate measurements revealed GDD fluctuations of 300-500 $fs^2$, which were significantly larger than the system uncertainty, suggesting that the existence of intrinsic pulse-to-pulse spectral phase fluctuations of the MIR pulses generated via IPDFG. The measured TOD fluctuations were approximately $2\times10^5$ $fs^3$, exceeding the TOD change introduced in the present experiments ($1.8\times10^4$ $fs^3$), which further supports the existence of intrinsic pulse-to-pulse spectral phase fluctuations. It is worthwhile to note that, like other spectral phase characterization techniques, TSUBAME does not measure the carrier-envelope phase (CEP). In the present implementation, however, the MIR pulses are inherently CEP-stable because they are generated through the IPDFG process.

The measurable MIR spectral range (4.98-5.30 μm) is currently limited by the MIR generation process rather than the detection scheme. Further optimization of the PPLN phase-matching condition or the use of alternative nonlinear crystals could expand the MIR spectral range to 3-16 μm, using 1 μm-pumped IPDFG[34]. With sufficiently broadband MIR generation, the short-wavelength detection limit is set by the spectral bandwidth of the broadband TSUBAME pulse, which must overlap with that of the upconverted pulse. In contrast, the long-wavelength side is not fundamentally constrained in the current scheme, as spectral overlap is inherently preserved between the upconverted pulse and the broadband TSUBAME pulse. At present, the 30-dB cutoff of the broadband TSUBAME spectrum is 810 nm, corresponding to an MIR limit of 3.8 μm. Extending detection toward shorter MIR wavelengths therefore requires further spectral broadening of the broadband TSUBAME pulse, achievable by increasing the input power or fiber length for SPM. For example, detecting 3 μm MIR pulse would require extension of the broadband TSUBAME to 768 nm. At this wavelength, the propagation loss[32] of SMF-28 (3.8 dB/km) remains moderate and does not impose a critical limitation. Therefore, the TSUBAME framework can, in principle, cover a much broader MIR spectral range than demonstrated here, provided that sufficient signal power is maintained.

The spectral resolution of TSUBAME is determined by the bandwidth of the narrowband TSUBAME pulse, the detector bandwidth, and the time-stretch fiber length. In our implementation, the narrowband TSUBAME pulse has an FWHM of 0.27 nm centered at 1032.8 nm. The employed PD has a bandwidth of 10 GHz, with an impulse response of 51 ps (FWHM). This impulse response corresponds to a response bandwidth of 0.15 nm when combined with 4-km SMF in our time-stretch detection. Therefore, the overall spectral resolution is estimated to be 0.24 nm at the detection wavelength (860.4 nm), corresponding to 8.7 nm (3.26 $cm^{-1}$) at the MIR wavelength (5.15 μm) before upconversion. The current spectral resolution is sufficient to resolve our MIR pulses with a $1/e^2$ bandwidth of 320 nm (121 $cm^{-1}$). Further improvement in spectral resolution can be achieved by reducing the narrowband TSUBAME bandwidth, using a PD with a larger bandwidth, and increasing the time-stretch fiber length.

**TSUBAME towards higher speed**

As a time-stretch-based technique, the speed of TSUBAME is fundamentally limited by the laser repetition rate due to its pulse-to-pulse measurement nature. Apart from the results presented here, we also verified TSUBAME operation at a rate of 50 MHz, the native repetition rate of our mode-locked laser. However, the spectral broadening system used to generate the broadband TSUBAME pulse was not sufficiently stable for long-term operation due to thermally induced fluctuations in fiber coupling into the short SMF. At 50 MHz, the high average power (3.5 W) required for sufficient spectral broadening caused heating at the fiber input facet, leading to coupling instability and even damage at the air-glass interface. Reducing the repetition rate to 1 MHz lowered the required average power (48.3 mW), enabling stable spectral broadening and reliable TSUBAME operation. Further improvement in measurement speed will require optimization of the spectral broadening system, for example, by using end-capped fibers to enhance stability[35]. On the detection side, the maximum measurement rate is achieved at a 100% duty cycle, where the interferometric stretched pulse duration matches the pulse repetition period. However, excessively high repetition rates can lead to insufficient spectral resolution. To overcome the trade-off between measurement rate and spectral resolution, incorporating compressive sensing techniques may enable TSUBAME operation at the GHz level[36]. In this case, we estimate that the required average power of MIR pulses for TSUBAME will be 40 mW, which should still be practically available.

**TSUBAME for arbitrary MIR pulsed sources**

In this work, we demonstrated TSUBAME using a Yb-doped fiber laser, with MIR pulses generated via IPDFG.

However, this is not the only possible implementation. Besides IPDFG, other common sources of ultrafast MIR pulses include optical parametric oscillators (OPOs)[37] and MIR mode-locked lasers[9]. For OPO-based systems, the narrowband and broadband TSUBAME pulses can be derived from the OPO pump in a manner similar to our implementation, particularly when Yb-based pump lasers are used. OPOs driven by other lasers may also be compatible, provided that a broadband TSUBAME pulse with sufficient bandwidth can be generated to overlap with the upconverted pulse. For MIR mode-locked lasers or other independent ultrafast MIR sources, it is preferable to use an independent TSUBAME source in the NIR, with its repetition rate synchronized to that of the MIR oscillator to ensure temporal overlap. More specifically, the timing jitter between the broadband TSUBAME pulse and the upconverted pulse should remain below the temporal resolution corresponding to the spectral resolution of the detection system, which is 97.24 GHz, or 10.3 ps in the time domain, in our setup. In addition, efficient upconversion requires that the relative timing jitter between the narrowband TSUBAME pulse and the MIR pulses be maintained at the sub-picosecond level. In practice, these requirements should be experimentally feasible by locking the repetition rates of the two pulse sources, given that sub-femtosecond timing jitter has already been demonstrated[38,39]. In contrast, CEP stabilization is unnecessary because CEP fluctuations appear only as a constant offset in the retrieved spectral phase and do not affect the phase analysis. TSUBAME sources based on Yb- or Er-doped fiber lasers offer flexible repetition-rate control. Considering detection efficiency, particularly the propagation loss in the time-stretch fiber, Er-doped fiber lasers may be advantageous for MIR wavelengths in the 3-6 μm range, as the corresponding upconversion wavelengths are longer and thus experience lower fiber propagation loss.

**Comparison with existing spectral-phase-resolved time-stretch approaches in other spectral regions**

Several approaches for spectral phase retrieval based on time-stretch techniques have previously been demonstrated in the NIR and THz regions. In the NIR region, spectral phase retrieval has been achieved by applying iterative reconstruction algorithms to time-stretched spectral intensity waveforms[40,41]. However, these approaches generally require prior knowledge of the input pulse intensity and may suffer from ambiguities and reconstruction errors inherent to iterative phase retrieval[42]. Time-stretch spectral interferometry has also been reported, including MHz-rate pulse-to-pulse characterization in the NIR region. However, phase retrieval in these demonstrations relied on trained neural networks[43]. In contrast, TSUBAME retrieves the spectral phase directly through a non-iterative calculation based on spectral interferometry, without requiring iterative optimization or machine-learning-based reconstruction.

Time-stretch EOS has enabled pulse-resolved measurements of THz waveforms at kHz acquisition rates[44,45]. Extending EOS-based time-stretch approaches to the MIR region, however, is fundamentally challenging because resolving MIR electric fields requires substantially higher temporal resolution. In contrast, TSUBAME measures the spectral phase rather than the electric-field waveform itself, thereby avoiding the stringent temporal-resolution requirements associated with direct field-resolved detection.

## Conclusion

In summary, we have introduced TSUBAME as a pulse-to-pulse spectral phase characterization technique for broadband MIR pulses operating at the laser repetition rate. We experimentally demonstrated the retrieval of spectral intensity and phase of MIR pulses over a range of 4.98-5.30 μm, with a spectral resolution of 8.7 nm and a measurement rate of 1 MHz. To the best of our knowledge, this represents the first demonstration of pulse-resolved MIR spectral phase characterization beyond the MHz rate, establishing a speed benchmark more than 3 orders of magnitude faster than existing methods. Although demonstrated at 1 MHz, the measurement speed is fundamentally limited only by the repetition rate of the MIR source, indicating strong potential for operating at significantly higher rates, potentially reaching the GHz regime. Despite operating in a single-shot manner, TSUBAME achieves a high SNR of up to 65 and a phase precision as high as 3.73 mrad, enabling practical applications of the technique. We further verified accurate measurements of phase variations in MIR pulses with a temporal resolution of 10 μs. Beyond the present demonstration, TSUBAME can be extended to broader MIR wavelength ranges and applied to a wide variety of ultrashort MIR sources. The developed technique opens new opportunities for investigating pulse-to-pulse MIR laser dynamics across a wide range of repetition rates, and for applications requiring precise spectral phase control and monitoring in strong-field physics and molecular sciences.

## Funding

This work was supported by Japan Society for the Promotion of Science (23H00273, 25H01386, T.I.), JST FOREST Program (JPMJFR236C, T.I.), RIKEN TRIP initiative (T.I.). Z.D. was supported by MERIT-WINGS Program from the University of Tokyo and JST SPRING (JPMJSP2108). This work has been co-funded by the European Union's Horizon Europe Research and Innovation Programme under agreement 101070700 (project MIRAQLS). Z.X. was supported by a Postdoctoral Research Fellowship for Young Scientists from the Japan Society for the Promotion of Science and a Postdoctoral Fellowship for Research in Japan from the Japan Society for the Promotion of Science. D.V.S. was supported by the Japan Society for the Promotion of Science Invitation Fellowship for Research in Japan (Short-term).

## Acknowledgements

The authors acknowledge Takuma Nakamura for providing system maintenance instructions and Makoto Shoshin for inspiring discussions and valuable comments on the manuscript.

## Author contributions

Z.D., D.V.S. and T.I. conceived the idea. Z.D., Z.X., K.H. and G.D. developed the experimental setup. Z.D. performed the experiments and analyzed the data. Z.X. validated the data analysis. Z.D., Z.X., K.H. and T.I. discussed the interpretation of the results. G.D. performed confirmatory experiments. Z.X., K.H. and T.I. supervised the work. Z.D., Z.X., K.H. and T.I. wrote the manuscript with inputs from other authors.

## Conflicts of Interest

The authors declare no conflicts of interest.

## Data availability

The data presented in the manuscript are available from the corresponding author upon reasonable request.

## References

1. Mitrofanov, A. V *et al.* Chirp-controlled high-harmonic and attosecond-pulse generation via coherent-wake plasma emission driven by mid-infrared laser pulses. *Opt. Lett.* **45**, 750–753 (2020).

2. Steinleitner, P. *et al.* Single-cycle infrared waveform control. *Nat. Photonics* **16**, 512–518 (2022).

3. Tsubouchi, M. & Momose, T. Pulse shaping and its characterization of mid-infrared femtosecond pulses:

Toward coherent control of molecules in the ground electronic states. *Opt. Commun.* **282**, 3757–3764 (2009).

4. Morichika, I., Murata, K., Sakurai, A., Ishii, K. & Ashihara, S. Molecular ground-state dissociation in the condensed phase employing plasmonic field enhancement of chirped mid-infrared pulses. *Nat. Commun.* **10**, 3893 (2019).

5. Thiré, N. *et al.* Highly stable, 15 W, few-cycle, 65 mrad CEP-noise mid-IR OPCPA for statistical physics. *Opt. Express* **26**, 26907–26915 (2018).

6. Konstantin L. Vodopyanov. Laser-based Mid-infrared Sources and Applications. in *Laser-based Mid-infrared Sources and Applications* 7–246 (2020). doi:https://doi.org/10.1002/9781119074557.fmatter.

7. Pires, H., Baudisch, M., Sanchez, D., Hemmer, M. & Biegert, J. Ultrashort pulse generation in the mid-IR. *Prog. Quantum Electron.* **43**, 1–30 (2015).

8. Huang, J., Pang, M., Jiang, X., He, W. & Russell, P. StJ. Route from single-pulse to multi-pulse states in a mid-infrared soliton fiber laser. *Opt. Express* **27**, 26392–26404 (2019).

9. Ma, J., Qin, Z., Xie, G., Qian, L. & Tang, D. Review of mid-infrared mode-locked laser sources in the 2.0 μm–3.5 μm spectral region. *Appl. Phys. Rev.* **6**, 021317 (2019).

10. Grebnev, K., Perminov, B., Fernandez, T. T., Fuerbach, A. & Chernysheva, M. Fluoride and chalcogenide glass fiber components for mid-infrared lasers and amplifiers: Breakthroughs, challenges, and future perspective. *APL Photonics* **9**, 110901 (2024).

11. Henderson-Sapir, O. & Ottaway, D. J. A decade of mid-infrared, 3.5 μm dual-wavelength pumped fiber lasers, review and perspective. *APL Photonics* **9**, 100901 (2024).

12. Riek, C. *et al.* Direct sampling of electric-field vacuum fluctuations. *Science (1979).* **350**, 420–423 (2015).

13. Chekhova, M. V, Leuchs, G. & Żukowski, M. Bright squeezed vacuum: Entanglement of macroscopic light beams. *Opt. Commun.* **337**, 27–43 (2015).

14. Virally, S., Cusson, P. & Seletskiy, D. V. Enhanced Electro-optic Sampling with Quantum Probes. *Phys. Rev. Lett.* **127**, 270504 (2021).

15. Lee, K. F., Kubarych, K. J., Bonvalet, A. & Joffre, M. Characterization of mid-infrared femtosecond pulses [Invited]. *Journal of the Optical Society of America B* **25**, A54–A62 (2008).

16. Li, Y. *et al.* Accurate characterization of mid-infrared ultrashort pulse based on second-harmonic-generation frequency-resolved optical gating. *Opt. Laser Technol.* **120**, 105671 (2019).

17. Bates, P. K., Chalus, O. & Biegert, J. Ultrashort pulse characterization in the mid-infrared. *Opt. Lett.* **35**, 1377–1379 (2010).

18. Nicolai, F., Müller, N., Manzoni, C., Cerullo, G. & Buckup, T. Acousto-optic modulator based dispersion

scan for phase characterization and shaping of femtosecond mid-infrared pulses. *Opt. Express* **29**, 20970–20980 (2021).

19. Geib, N. C. *et al.* Discrete dispersion scan setup for measuring few-cycle laser pulses in the mid-infrared. *Opt. Lett.* **45**, 5295–5298 (2020).

20. Kubarych, K. J., Joffre, M., Moore, A., Belabas, N. & Jonas, D. M. Mid-infrared electric field characterization using a visible charge-coupled-device-based spectrometer. *Opt. Lett.* **30**, 1228–1230 (2005).

21. Kugel, T., Okazaki, D., Arai, K. & Ashihara, S. Direct electric-field reconstruction of few-cycle mid-infrared pulses in the nanojoule energy range. *Appl. Opt.* **61**, 1076–1081 (2022).

22. Kempf, H. *et al.* Direct sampling of femtosecond electric-field waveforms from an optical parametric oscillator. *APL Photonics* **9**, 036111 (2024).

23. Weigel, A. *et al.* Ultra-rapid electro-optic sampling of octave-spanning mid-infrared waveforms. *Opt. Express* **29**, 20747–20764 (2021).

24. Benea-Chelmus, I.-C. *et al.* Electro-optic sampling of classical and quantum light. *Optica* **12**, 546–563 (2025).

25. Mahjoubfar, A. *et al.* Time stretch and its applications. *Nat. Photonics* **11**, 341–351 (2017).

26. Godin, T. *et al.* Recent advances on time-stretch dispersive Fourier transform and its applications. *Adv. Phys. X* **7**, 2067487 (2022).

27. Kawai, A. *et al.* Time-stretch infrared spectroscopy. *Commun. Phys.* **3**, 152 (2020).

28. Hashimoto, K. *et al.* Upconversion time-stretch infrared spectroscopy. *Light Sci. Appl.* **12**, 48 (2023).

29. He, L. *et al.* Long-wavelength infrared upconversion time-stretch spectroscopy. *Appl. Phys. Lett.* **125**, 071102 (2024).

30. Nakamura, T., Ramaiah Badarla, V., Hashimoto, K., Schunemann, P. G. & Ideguchi, T. Simple approach to broadband mid-infrared pulse generation with a mode-locked Yb-doped fiber laser. *Opt. Lett.* **47**, 1790–1793 (2022).

31. Boilard, T., Vallée, R. & Bernier, M. Probing the dispersive properties of optical fibers with an array of femtosecond-written fiber Bragg gratings. *Sci. Rep.* **12**, 4350 (2022).

32. Hill, C. *et al.* Monolithic All-Semiconductor PCSELs emitting at 1.3µm. in *2021 27th International Semiconductor Laser Conference (ISLC)* 1–2 (2021).

33. Malitson, I. H. A Redetermination of Some Optical Properties of Calcium Fluoride. *Appl. Opt.* **2**, 1103–1107 (1963).

34. Zhang, J. *et al.* Intra-pulse difference-frequency generation of mid-infrared (2.7-20μm) by random quasi-phase-matching. *Opt. Lett.* **44**, 2986–2989 (2019).

35. Vesco, G., Moretti, L., Della Chiesa, G., Marangoni, M. & Gatti, D. Broadband milliwatt-level pulse train beyond 6 μm from a femtosecond Yb oscillator. *Optics Continuum* **4**, 1621–1627 (2025).

36. Kawai, A., Horisaki, R. & Ideguchi, T. Compressive time-stretch spectroscopy with pulse-by-pulse intensity modulation. *Opt. Lett.* **49**, 3468–3471 (2024).

37. Vainio, M. & Halonen, L. Mid-infrared optical parametric oscillators and frequency combs for molecular spectroscopy. *Physical Chemistry Chemical Physics* **18**, 4266–4294 (2016).

38. Hillbrand, J., Andrews, A. M., Detz, H., Strasser, G. & Schwarz, B. Coherent injection locking of quantum cascade laser frequency combs. *Nat. Photonics* **13**, 101–104 (2019).

39. Vasilyev, S. *et al.* Middle-IR frequency comb based on Cr:ZnS laser. *Opt. Express* **27**, 35079–35087 (2019).

40. Solli, D. R., Gupta, S. & Jalali, B. Optical phase recovery in the dispersive Fourier transform. *Appl. Phys. Lett.* **95**, 231108 (2009).

41. Ryczkowski, P. *et al.* Real-time full-field characterization of transient dissipative soliton dynamics in a mode-locked laser. *Nat. Photonics* **12**, 221–227 (2018).

42. Xu, Y., Ren, Z., Wong, K. K. Y. & Tsia, K. Overcoming the limitation of phase retrieval using Gerchberg-Saxton-like algorithm in optical fiber time-stretch systems. *Opt. Lett.* **40**, 3595–3598 (2015).

43. Pu, G., Luo, C., Hu, W. & Yi, L. Intelligent single-shot full-field characterization over femtosecond pulses. *Nat. Commun.* **16**, 11621 (2025).

44. Szwaj, C. *et al.* High sensitivity photonic time-stretch electro-optic sampling of terahertz pulses. *Review of Scientific Instruments* **87**, 103111 (2016).

45. Couture, N. *et al.* Single-pulse terahertz spectroscopy monitoring sub-millisecond time dynamics at a rate of 50 kHz. *Nat. Commun.* **14**, 2595 (2023).